\documentclass[journal,twoside,web]{ieeecolor}
\usepackage{tuffc}
\usepackage{cite}
\usepackage{algorithmic}
\usepackage{graphicx}
\usepackage{textcomp}
\usepackage{wrapfig,colortbl}
\usepackage{multirow}
\usepackage{amsmath,amsfonts,amssymb}
\usepackage{mathrsfs}
\usepackage{newtxmath}
\usepackage{booktabs}
\usepackage{hyperref} 
\usepackage{algorithmic}
\usepackage{graphicx}
\usepackage{subfigure}
\usepackage{textcomp}
\usepackage{pbalance}
\definecolor{abstractbg}{rgb}{1,0.969,0.914}
\def\BibTeX{{\rm B\kern-.05em{\sc i\kern-.025em b}\kern-.08em
    T\kern-.1667em\lower.7ex\hbox{E}\kern-.125emX}}
\begin{document}
\title{Ultrafast Cardiac Imaging Using Deep Learning For Speckle-Tracking Echocardiography}
\author{
Jingfeng Lu,
        Fabien Millioz,
        François Varray,
        Jonathan Porée,
        Jean Provost,\\
        Olivier Bernard,
        Damien Garcia,
        and Denis Friboulet
\thanks{This work was supported in part by Sichuan University Postdoctoral
Interdisciplinary Innovation Fund under Grant JCXK2232, and partially supported by the framework of the LABEX PRIMES (ANR-11-LABX-0063) of Université de Lyon and the LABEX CELYA (ANR-11-LABX-0060) of Université de Lyon, both within the program "Investissements d'Avenir" (ANR-11-IDEX-0007) operated by the French National Research Agency (ANR). The \textit{in vivo} part was partially supported by a Tremplin International - INSERM grant (DELEA-2022-480/DELCP-2022-271)}
\thanks{J. Lu is affiliated with the School of Cyber Science and Engineering, Sichuan University, Chengdu, China, he was with the School of Instrumentation Science and Engineering, Harbin Institute of Technology, Harbin, China at the time of writing this paper.}
\thanks{F. Millioz, F. Varray, O. Bernard, D. Garcia, and D. Friboulet are affiliated with the University of Lyon, CREATIS, CNRS UMR 5220, Inserm U1294, INSA-Lyon, University of Lyon 1, Villeurbanne, France.}
\thanks{J. Porée and J. Provost are affiliated with Engineering Physics Department, Polytechnique Montréal, Montréal, Canada.}
\thanks{The proposed model with trained weights in Open Neural Network Exchange (ONNX) format and a demo script are available at \href{https://github.com/Jingfeng-LU/Deep-Ultrafast-Echocardiography}{https://github.com/Jingfeng-LU/Deep-Ultrafast-Echocardiography}}
}

\IEEEtitleabstractindextext{%
\fcolorbox{abstractbg}{abstractbg}{%
\begin{minipage}{\textwidth}\rightskip2em\leftskip\rightskip\bigskip
\begin{wrapfigure}[31]{r}{2.8in}%
\hspace{-3pc}\includegraphics[width=2.8in]{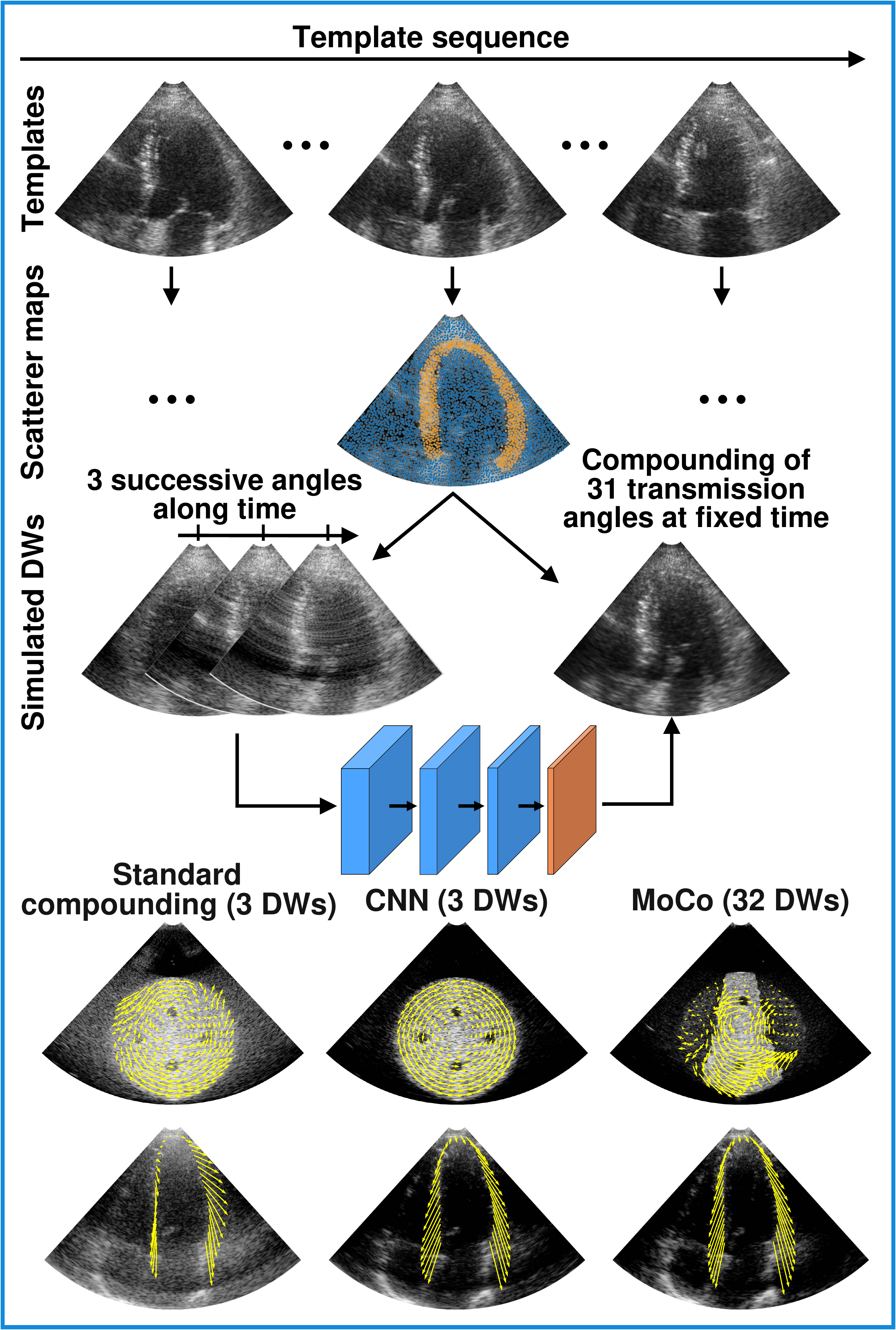}
\end{wrapfigure}%
\begin{abstract}
High-quality ultrafast ultrasound imaging is based on coherent compounding from multiple transmissions of plane waves (PW) or diverging waves (DW). However, compounding results in reduced frame rate, as well as destructive interferences from high-velocity tissue motion if motion compensation (MoCo) is not considered. While many studies have recently shown the interest of deep learning for the reconstruction of high-quality static images from PW or DW, its ability to achieve such performance while maintaining the capability of tracking cardiac motion has yet to be assessed. In this paper, we addressed such issue by deploying a complex-weighted convolutional neural network (CNN) for image reconstruction and a state-of-the-art speckle tracking method. The evaluation of this approach was first performed by designing an adapted simulation framework, which provides specific reference data, i.e. high quality, motion artifact-free cardiac images. The obtained results showed that, while using only three DWs as input, the CNN-based approach yielded an image quality and a motion accuracy equivalent to those obtained by compounding 31 DWs free of motion artifacts. The performance was then further evaluated on non-simulated, experimental \textit{in vitro} data, using a spinning disk phantom. This experiment demonstrated that our approach yielded high-quality image reconstruction and motion estimation, under a large range of velocities and outperforms a state-of-the-art MoCo-based approach at high velocities. 
Our method was finally assessed on \textit{in vivo} datasets and showed consistent improvement in image quality and motion estimation compared to standard compounding. This demonstrates the feasibility and effectiveness of deep learning reconstruction for ultrafast speckle-tracking echocardiography. 
\end{abstract}

\bigskip
\end{minipage}}}

\maketitle

\begin{table*}[!t]
\arrayrulecolor{subsectioncolor}
\setlength{\arrayrulewidth}{1pt}
{\sffamily\bfseries\begin{tabular}{lp{6.75in}}\hline
\rowcolor{abstractbg}\multicolumn{2}{l}{\color{subsectioncolor}{\itshape
Highlights}{\Huge\strut}}\\
\rowcolor{abstractbg}$\bullet$ & A novel framework for deep learning-based ultrafast ultrasound cardiac imaging, and evaluation of its temporal coherence for speckle-tracking echocardiography.\\
\rowcolor{abstractbg}$\bullet${\large\strut} &The proposed approach demonstrates high image quality while preserving consistent speckle pattern for tracking high-speed tissue motion, when applied on simulated, \textit{in-vitro}, and \textit{in-vivo} data. \\

\rowcolor{abstractbg}$\bullet${\large\strut} & This approach has the potential to improve cardiac diagnosis using dynamic analysis at ultra-high frame rates (1500 frames/second in our experiments), owing to its high imaging quality and detectable tissue speed.
 \\[2em]\hline
\end{tabular}}
\setlength{\arrayrulewidth}{0.4pt}
\arrayrulecolor{black}
\end{table*}

\section{Introduction}
\label{sec:introduction}
\IEEEPARstart{C}{onventional} echocardiography typically reaches 30 to 100 frames/second with a line-by-line ultrasound system \cite{cikes2014ultrafast}, where multiple sectors of the entire field of view are insonified using sequential narrow beams. Such a frame rate is sufficient for speckle tracking echocardiography in a resting adult whose heartbeat is approximately 70 per minute \cite{voigt2015definitions}. However, higher frame rates may be required in some imaging situations, e.g. i) stress echocardiography \cite{pellikka2020guidelines}, ii) all-four-chamber strain imaging \cite{addetia2016simultaneous,joos2018high}, iii) electromechanical wave imaging \cite{provost2009electromechanical}, and iv) myocardial shear wave imaging \cite{villemain2019myocardial,salles20213d}. 

To achieve higher frame rates, wide beam imaging systems using plane wave (PW) \cite{sandrin1999time} or diverging wave (DW) \cite{hasegawa2011high} have been developed to reconstruct the full field of view in a single transmit. Such schemes can provide frame rates up to several thousands of frames/second. Ultrafast imaging can thus capture short-duration events in cardiac cycles, potentially improving the accuracy of motion tracking. An increase in frame rate, however, comes at the expense of a degraded resolution and contrast compared with conventional focused imaging. An approach to improve the quality of ultrafast imaging is coherent compounding \cite{montaldo2009coherent,papadacci2014high}, where consecutive steered transmit beams at different angles are transmitted and the backscattered echoes are summed coherently. Coherent compounding can yield images of high quality, at the expense of the frame rate compared with single plane-wave/diverging-wave imaging. In practice, a trade-off must be made between image quality and frame rate.

In cardiac imaging, which involves fast-moving tissues, motion artifacts may ensue during a large number of transmit-receive events for coherent compounding, which in turn degrades the quality of compounded images. As a result, inter-acquisition motion must be considered during the compounding process to mitigate image degradation. Motion compensation (MoCo) methods have thus been introduced to provide high-quality echocardiography \cite{poree2016high}, while preserving the speckle patterns for carrying out efficient speckle tracking \cite{poree2018dual}. A typical frame rate of MoCo-based compounding approaches for high-quality echocardiography in adults is around 250 frames/second, and is still limited by the temporal-spatial resolution trade-off. Yet, MoCo techniques can be impaired by the limited quality of the reference frame in registration-based MoCo \cite{stanziola2018motion}, or by aliasing due to the Nyquist limit in Doppler-based MoCo \cite{poree2016high}. Exploiting the full potential of ultrafast cardiac imaging in terms of frame rate and image quality is a very active area of interest.

Recently, a wide variety of studies based on deep learning have sought to improve the image quality of high-frame-rate ultrasound imaging, such as PW imaging\cite{gasse2017high, nair2018deep, lu2018unsupervised,khan2019deep, zhou2019ultrafast,perdios2020cnn, zhang2021ultrasound,lu2022improving}, DW imaging\cite{ghani2019high, lu2020reconstruction,lu2022complex}, synthetic transmit aperture imaging\cite{chen2021apodnet,zhang2022deep}, multiline acquisition\cite{senouf2018high, vedula2019learning}, and multiline transmission\cite{vedula2018high}. 
Among these works, the studies in\cite{senouf2018high,vedula2019learning,vedula2018high} were devoted to cardiac imaging. In these studies, the image quality was typically only evaluated on static cardiac frames in terms of image contrast, as well as similarity to high-quality references from single-line acquisitions. 
However, the interest of high-frame-rate echocardiography lies in deciphering highly transient physiological events using cardiac sequences in high temporal resolution. 
Deep learning-based methods have yet to show that it is possible to achieve high imaging quality while maintaining the capability of tracking blood and tissue motion.

However, very few studies have investigated the properties of deep learning-based reconstruction for motion estimation. One notable exception is the study in \cite{perdios2020cnn}, which performed such investigation in the context of vascular imaging using PW imaging. In this work, motion estimation assessment was performed from numerical simulations. The presented \textit{in vivo} results are obtained from one proof-of-concept case acquired using a low frame rate (10 Hz) and concerned displacements of slow-moving tissue between the skin and the carotid (i.e. 1.25 mm/s), which are thus far from the much more complex and fast cardiac motion patterns. 

DW transmissions are particularly interesting in echocardiography, since they can cover a wide and deep cardiac view (e.g. a four-chamber view) with a single transmission \cite{joos2018high,poree2018dual}. However, very few studies \cite{lu2019fast,ghani2019high,lu2020complex} are devoted to deep learning-based processing of
DW. We previously described a deep learning framework for high-quality ultrafast DW imaging reconstruction\cite{lu2020reconstruction,lu2022complex}. In \cite{lu2022complex}, a complex-values CNN was exploited for reconstructing high-quality ultrasound images from DAS beamformed I/Q signals acquired with a small number (typically three) of DW emissions.

In this paper, we propose a deep learning-based reconstruction approach for ultrafast cardiac imaging, and showed that it yielded high imaging quality while achieving a high accuracy in tracking tissue motion.

\begin{figure*}[!t]
\centering
\includegraphics[width=1.9\columnwidth, angle=0]{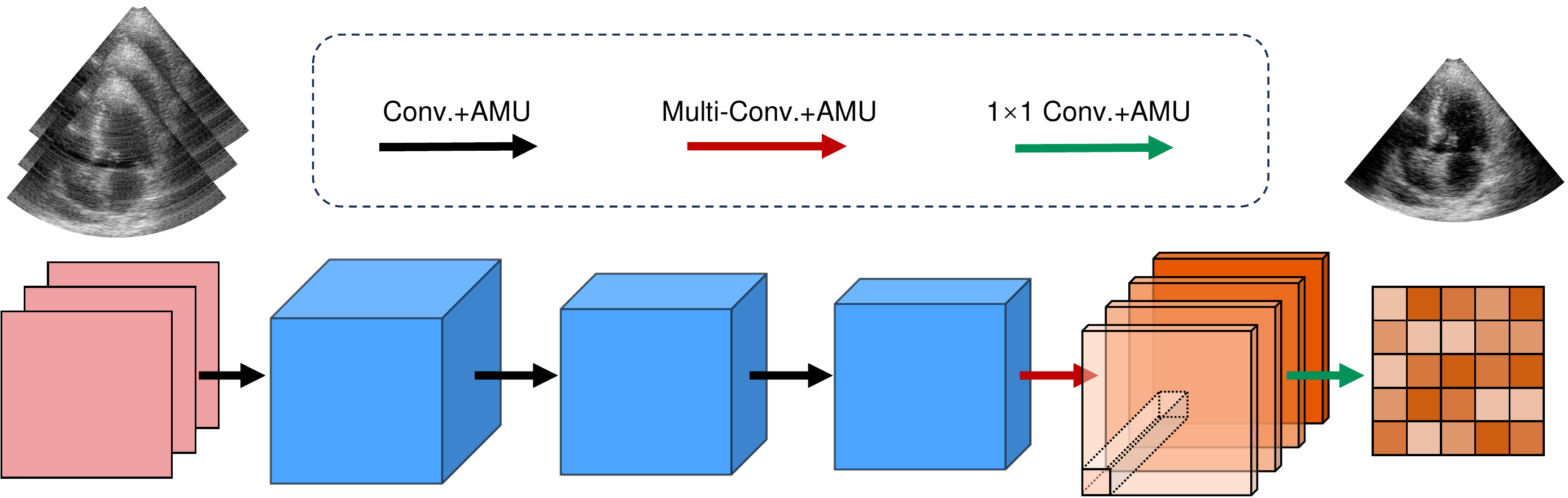}
\caption{Block diagram of the model architecture}
\label{network_diagram}
\end{figure*}

The main contributions of this work are the following:

1) We designed a specific numerical simulation pipeline providing clinical-like ultrasound cardiac sequences, for proper training of deep learning models and assessing reconstruction and motion estimation quality. In particular, this simulation provided task-specific reference data, by allowing imaging the heart as if it was frozen in time, yielding high-quality, motion artifact-free images from the compounding of 31 DWs.

2) This simulation framework allowed to demonstrate that the proposed method reconstructs high-quality cardiac images using only 3 DWs, yielding an image quality and a motion accuracy equivalent to those obtained from the reference data, i.e. from the compounding of 31 motion artifact-free DWs. 

3) The CNN trained on the cardiac simulation and the resulting motion estimation were further evaluated on non-simulated, experimental \textit{in vitro} data, using a spinning disk phantom. This experiment allowed to demonstrate that our approach yields high-quality image reconstruction and accurate motion estimation under a large range of velocities. In particular, we showed that the proposed approach outperforms a state-of-the-art MoCo-based approach at high velocities.

4) The approach was assessed from \textit{in vivo} cardiac acquisitions and the results showed consistent improvement in imaging quality and motion estimation, as compared with standard compounding, which demonstrated the feasibility and effectiveness of the proposed approach for ultrafast speckle-tracking echocardiography.

The remainder of this article is organized as follows. In Section II, we present our framework for ultrafast DW imaging, data simulation, and speckle tracking. The experimental configurations for realistic imaging and model training are presented in Section III. Section IV presents the experimental results, which are further discussed in Section V. Finally, our concluding remarks are given in Section VI.

\begin{table}[!t]
\renewcommand{\arraystretch}{1.3}
\setlength{\tabcolsep}{3.2mm}
\caption{Architecture of the proposed network}
\label{network}
\centering
\begin{tabular}{c c c c c}
\toprule
Feature size & Kernel size & Kernel number& Activation\\ 
\midrule
$m \times h \times w$&–&–&–\\
$64 \times h \times w$ & $~3 \times 3~$  & 256 & 4-piece AMU\\
$32 \times h \times w$ & $~5 \times 5~$ & 128 & 4-piece AMU\\
$16 \times h \times w$ & $~9 \times 9~$ & 64 & 4-piece AMU\\
\cline{2-4}
\multirow{4} * {$8 \times h \times w$} & $11 \times 11$ & 8 & 4-piece AMU\\
~ & $13 \times 13$ & 8 & 4-piece AMU\\
~ & $15 \times 15$ & 8 & 4-piece AMU\\
~ & $17 \times 17$ & 8 & 4-piece AMU\\
\cline{2-4}
 $1 \times h \times w$ & $1 \times 1$ & 4 & 4-piece AMU\\
\bottomrule
\end{tabular}
\end{table}

\section{Methods}
\label{sec:method}

\subsection{Deep Learning-based DW imaging}

We consider a small number of $m$ steered DW transmissions, each yielding raw I/Q image $\boldsymbol{\chi}_i$, $i\in [1, m]$. A DAS beamformer was first applied to produce beamformed I/Q data $\boldsymbol{x}_i=\mathrm{DAS}(\boldsymbol{\chi} _i)$ of size $h \times w$, which represents the dimension of beamforming grid. 
The second step consists of reconstructing a high-quality I/Q image from the set of beamformed data $\boldsymbol{x}=\{ \boldsymbol{x_i}\}$. Instead of the standard compounding that requires a large number of transmits (typically on the order of thirty in cardiac imaging\cite{poree2016high}), we adopted a CNN-based model with learnable parameters $\boldsymbol{\theta}$ to seek an optimal DW compounding $\boldsymbol{\hat y} = \boldsymbol{f_\theta}(\boldsymbol{x})$ using supervised learning, with respect to high-quality targets $\boldsymbol y$. The references in this study were obtained from the compounding of a large number of motion artifact-free DWs. The model was trained by searching for the optimal convolutional weights $\hat{\boldsymbol{\theta}}$ with loss minimization,
\begin{equation}
\hat{\boldsymbol{\theta}}=\arg \min  \sum_{(\boldsymbol{x}, \boldsymbol{y})}^{{\boldsymbol{\mathcal{D}}}} \lVert \boldsymbol{f_ \theta}(\boldsymbol{x})-\boldsymbol{y}\rVert^2_2
\end{equation}
where $\boldsymbol{\mathcal{D}}$ is a dataset containing representative $(\boldsymbol{x},\boldsymbol{y})$ samples. 
In order to consistently process the I/Q data in the complex-valued domain, we adopted the complex-valued CNN model proposed in \cite{lu2022complex}, briefly described hereafter. The interested reader will find in \cite{lu2020reconstruction} and \cite{lu2022complex} more details about the justification and properties of this architecture.

Fig. \ref{network_diagram} shows the block diagram of the model architecture and Table \ref{network} provides a more detailed specification. As we use 3 DWs as input to the network, the input dimension $m$ is fixed to 3. The model is a fully convolutional network with five complex-valued convolution layers and amplitude maxout units (AMUs). We excluded spatial pooling for preserving the spatial information at the same resolution throughout the network. Each convolution layer is followed by an AMU activation, which activates elements corresponding to the element-wise maximum amplitude across several affine feature maps.

The second-to-last layer consists of the concatenation of multi-scale convolution kernels. $1 \times 1$ convolution is used in the last layer for channel dimension reduction and producing reconstructed I/Q images. 
This layer is equivalent to cross-channel parametric pooling, which can be trained to select element-wise the main feature that contributes to the production of the output element. 
As demonstrated in \cite{lu2020reconstruction}, the multi-scale convolution used in conjunction with $1 \times 1$ convolution and maxout activation allowed position-dependent features from different receptive fields to be captured, which helped to address the specific geometry of data sampling for DW imaging.

\subsection{Cardiac Data Simulation for Deep Learning}\label{sec_II_B}

Effective learning of the reconstruction model generally requires the dataset $\mathcal{D}$ to be representative of the joint distribution of input-output space, i.e. paired $(\boldsymbol{x,y})$ with a wide variety of cardiac geometries in the context of echocardiography. 
It is however challenging to obtain high-quality target images $\boldsymbol y$ with realistic acquisitions, since cardiac motion induces artifacts when a large number of transmit-receive events are used for coherent compounding. 
We therefore propose a simulation pipeline for generating clinical-like ultrasound cardiac sequences to tackle this issue. 
We briefly describe in the following the specificities of the simulation pipeline yielding the time-matched pairs $(\boldsymbol{x},\boldsymbol{y})$. The interested readers will find a more detailed description in \cite{alessandrini2015pipeline,evain2022motion}.

\subsubsection{Pipeline overview}

\begin{figure}[!t]
\centering
\includegraphics[width=0.9\columnwidth, angle=0]{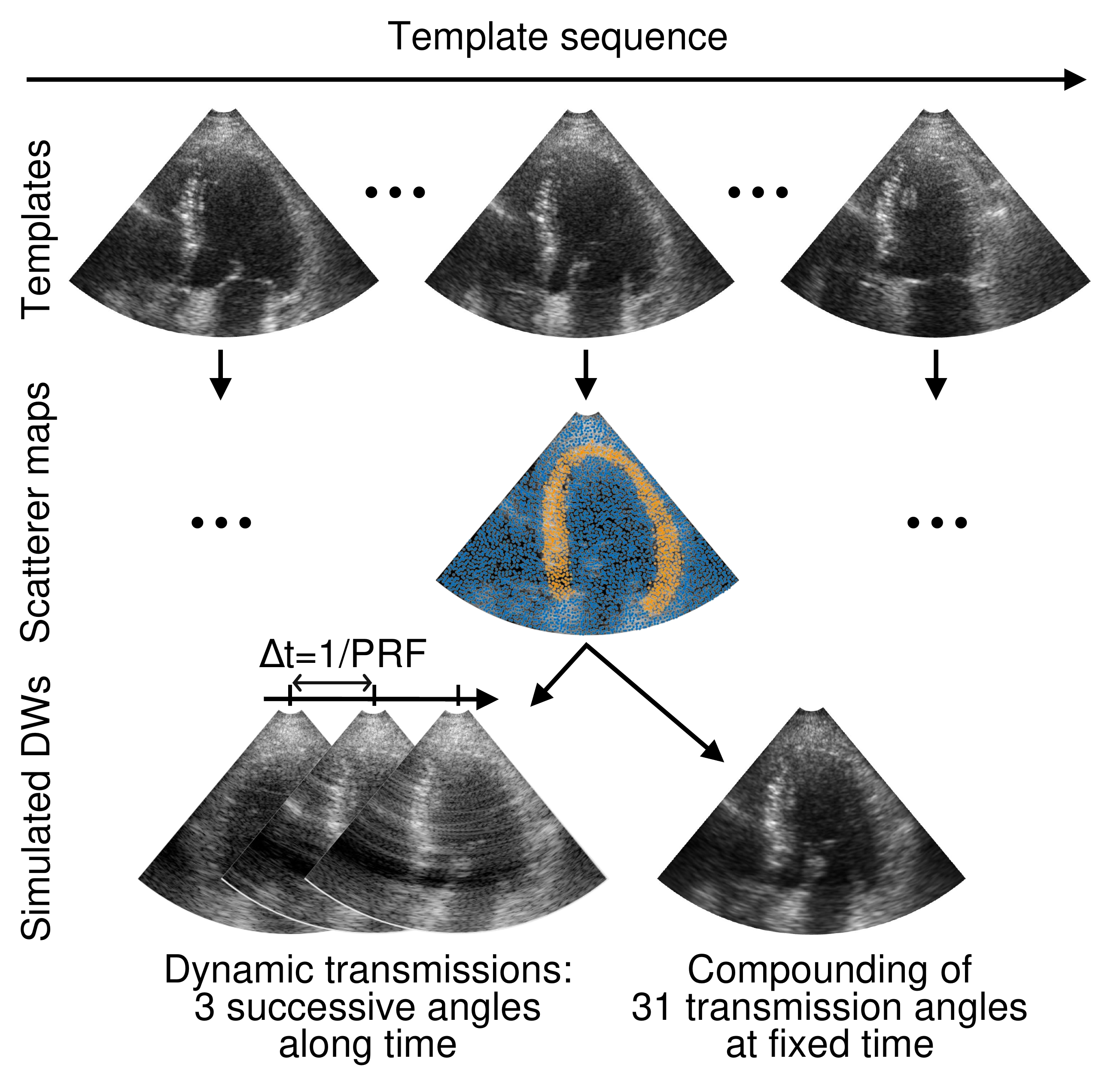}
\caption{Framework of cardiac imaging simulation. Based on real sequences, 3 DW transmissions were simulated with cardiac motion, and 31 DW transmissions were simulated at the frozen time.}
\label{simulation}
\end{figure}

The overall framework for the numerical simulation is illustrated in Fig. \ref{simulation}.
The pipeline starts with B-mode cardiac sequences from clinical recordings. These real images were used as the templates for cardiac geometries, which define the acoustic scatterer maps in the simulated field of view. Steered DW transmissions in the scatterer medium were simulated using a homemade open-source software called SIMUS \cite{garcia2022simus,cigier2022simus} for the generation of synthetic ultrasound data. To obtain motion artifact-free reference, each $\boldsymbol y$ was generated with steered DWs fired at a single time point (i.e. with a fixed scatterer map corresponding to the simulated frame), thus allowing imaging the heart as if it was frozen in time. On the opposite, $\boldsymbol x$ was produced in a dynamic manner for mimicking real acquisitions. The steered DW transmissions were simulated while taking into account the cardiac motion between each acquisition. The cardiac motion was simulated by the scatterer displacement using the strategy described at the end of this section.

\subsubsection{Cardiac geometry}

A carefully-defined medium of acoustic scatterers was exploited for the simulation of cardiac geometry. In order to obtain realistic speckle statistics, we followed the approach previously described in\cite{alessandrini2015pipeline}. For each template sequence, the scatterers were randomly distributed within the imaging sector at a density of 10 per square wavelength and the reflection coefficients $\mathrm{RC}_m$ of the scatterers were computed according to the local intensities $I_m$ of the B-mode templates as \cite{alessandrini2015pipeline}
\begin{equation}
\mathrm{RC}_{m}=\left(I_{m} / 255\right)^{(1 / \gamma)} \cdot \mathcal{N}(0,1)
\end{equation}
where $ \mathcal{N}(\cdot)$ is the normal distribution, and $\gamma$ is a constant for gamma compression.  As shown in \cite{alessandrini2015pipeline}, this strategy yields fully developed speckle for blood regions (i.e. Rayleigh statistics) and partially developed speckle in myocardial regions (i.e. sub-Rayleigh statistics). 
The initial myocardium scatterers were distributed in the manually annotated myocardium region of the first frame. The scatterer positions for the subsequent frames were computed using the strategy described in Section II-B-3. The reflection coefficients of myocardium scatterers were set constant to maintain the speckle texture throughout the simulated sequences. 

\subsubsection{Myocardium motion}

The simulation pipeline includes the motion fields in myocardium regions. The myocardium regions were obtained by manually delineating the endocardial and epicardial contours on the template frames. The regions of interest (ROIs) were then resampled to obtain time-varying surface meshes throughout the cardiac cycles. The myocardium motion fields were obtained with the propagation of the mesh over the full sequence. Such a scheme allows obtaining paired $(\boldsymbol{x},\boldsymbol{y})$ at any time point throughout the temporal trajectory of the simulated motion. The motion fields in the simulation were used as the ground truth for the evaluation of speckle tracking among consecutive reconstructed frames.

\subsection{Speckle tracking}

The tissue interframe motion was estimated on reconstructed frames using a speckle tracking approach described in \cite{perrot2018back}. 
This approach is based on a block-matching algorithm, where the normalized cross-correlation (NCC) is computed via a FFT for efficiency \cite{garcia2018introduction}. Let us note that other efficient approaches for NCC computation are possible\cite{luo2010fast}.
The approach can be summarized by the following process. The consecutive real-envelope images were divided into overlapping subwindows. The normalized cross-correlation (NCC) corresponding to each subwindow was computed on the subwindow pairs of two successive frames,  and the relative displacement within each subwindow is given by the location of the peak in each NCC. A parabolic peak fitting around the NCC peak was used to obtain the displacement with subpixel accuracy. 
To detect a wide range of motions, the process was iteratively performed in a coarse-to-fine manner by iteratively decreasing the size of the subwindows to refine the motion fields. 

\section{Experiments}

\subsection{Data acquisition}
\subsubsection{Simulated data}

We performed the cardiac sequence simulation using the pipeline described in Section II. The template sequences were extracted from the CAMUS open-access dataset \cite{leclerc2019deep} composed of exams from 500 patients acquired in clinical routine from the University Hospital of St-Etienne (France) within the regulation set by the local ethical committee of the hospital. 
A subset of 94 sequences with apical four-chamber views were selected for the simulation, resulting in 4324 frame templates with the corresponding cardiac texture and myocardial displacement fields. 

We simulated steered DW transmissions under the imaging settings corresponding to an ATL P4-2 probe: a 64-element phased array of 0.3-mm pitch and 50-$\mathrm{\mu}$m kerf. The center frequency was 3 MHz and the bandwidth was 2–4 MHz. For each template frame, we simulated 31 steered DW transmissions with tilt angles evenly spaced between $\pm 20^{\circ}$. The scatterer map was set fixed to obtain motion artifact-free references $\boldsymbol y$. We then simulated 3 steered DW transmissions $\boldsymbol x$ while taking into account the cardiac motion between each transmission. The steering angles were $-20^{\circ}$, $0^{\circ}$, and $20^{\circ}$. 
The motion was obtained by interpolating the reference motion field considering the pulse reception frequency (PRF), which was set to 3850 Hz in the simulation (i.e. $\Delta t = 0.26$ ms).
 Such a PRF allows round-trip ultrasonic transmissions of 20-cm depth between each acquisition, which is the largest axial depth of the template sequences.  
The synthetic RF data were sampled at 12 MHz, then demodulated and beamformed using a delay-and-sum (DAS) \cite{perrot2021so} to produce the beamformed I/Q data. 

\subsubsection{In vitro data}
The \textit{in vitro} data were acquired using a research scanner (Vantage 256, Verasonics Inc.) and an ATL P4-2 probe with the same probe settings used for the simulation and a PRF of 4500 Hz. Imaging was carried out on a 10-cm-diameter tissue-mimicking disk phantom with four anechoic cysts. The disk was mounted on a step motor and rotated at controlled speeds. The angular velocities ranged from 0 to 12 radians per second with an increment of one, giving a maximum speed of 60 cm/s at the disk periphery that encompasses the myocardial maximum velocity, i.e $\sim\,$15 cm/s for normal subjects \cite{garcia1996differentiation} to $\sim\,$30 cm/s for endurance athletes \cite{caso2000pulsed}. 
We performed two separate DW acquisition series for each rotational speed: the 3-angle sequence and the MoCo-based one. The MoCo-based sequences used a triangle steering strategy with 32-angle series and 50\% overlap \cite{poree2016high}. This yields a frame rate of 280 frames/second for MoCo and 1500 frames/second for the 3-angle sequence.
For each acquisition, the received raw RF data were sampled and processed with the same settings used for the simulation.

\subsubsection{In vivo data}
The \textit{in vivo} study was performed from five normal subjects and was approved by the Polytechnique Montréal Comité d’éthique et de la recherche (CER-2122-54-D).
Apical four-chamber views were examined. Two types of transmission, i.e. 3-angle sequences (used for CNN reconstruction) and 32-angle triangular sequences (used for MoCo-based reconstruction), were obtained with the same imaging and processing configurations used for the \textit{in vitro} experiment. 
During acquisition, these two transmission types were interleaved, in order to minimize the time interval between the corresponding reconstructions and make them comparable. Using a PRF of 4500 Hz and considering a maximum myocardial velocity of 15 cm/s in normal subjects, this indeed corresponds to a maximum displacement smaller than 0.07 mm, i.e. less than $\lambda$/7 (where $\lambda$ denote the wavelength using a central frequency of 3 MHz and a sound speed of 1540 m/s).

\subsection{Training implementation}

From the 94 simulated cases, 56 cases were randomly selected for training, and 19 cases were used as an independent validation set during training, to avoid overfitting. The remaining 19 cases were used for testing.
In the training stage, the model weights were initialized with the Xavier initializer \cite{glorot2010understand} and optimized by minimizing the training loss with mini-batch gradient descent. The batch size was set to 16 and the Adam\cite{kingma2014adam} optimizer was employed for gradient descent with back-propagation. The initial learning rate was set to $1 \times 10^{-4}$ and an early stopping strategy was used to adjust the learning rate. 
The learning rate was halved if the validation loss had not reduced for 10 epochs and the training ended if there had been 20 epochs without any validation loss reduction. 
The model was implemented using Pytorch library on an NVIDIA Tesla V100 GPU with 32 GB of memory, resulting in training time of about one day.

\subsection{Comparison Method}
For numerical experiments, the results obtained from the motion artifact-free sequences (i.e. compounding of 31 DWs simulated at frozen time) were considered as the reference. The CNN-based results were then compared with this reference and with the results obtained from standard compounding of 3 simulated DWs (which took into account the cardiac motion between each transmission.

For \textit{in vitro} and \textit{in vivo} experiments, as a motion artifact-free reference is not available, the results obtained from a state-of-the-art MoCo-based compounding \cite{poree2016high} were used as a reference. The CNN-based results were then compared with this reference and with the results obtained from standard compounding of 3 DWs.

\subsection{Evaluation Metrics}
\subsubsection{Image quality}
We evaluated the image quality using peak signal-to-noise ratio (PSNR) and structural similarity index (SSIM) in the numerical experiment. These two metrics measure the image similarity between the reconstructed images $\boldsymbol{\hat y}$ and the high-quality references $\boldsymbol{y}$.
\begin{equation}
\label{psnr}
{\rm PSNR} = 20 \log_{10}\frac{{\rm MAX}(\boldsymbol{y})}{\sqrt{\frac{1}{n}\| \hat {\boldsymbol{y}} - \boldsymbol{y} \|_2^2}}
\end{equation}
where $n$ is the number of pixels in the images.
\begin{equation}
\label{ssim}
{\rm SSIM} = \frac{(2\mu _{\hat {\boldsymbol{y}}} \mu _{\boldsymbol{y}} + C_1)(2\sigma _{\hat{\boldsymbol{y}}{\boldsymbol{y}}} + C_2)}{(\mu _{\hat {\boldsymbol{y}}}^2 + \mu _{\boldsymbol{y}}^2 + C_1)(\sigma _{\hat {\boldsymbol{y}}}^2 + \sigma _{\boldsymbol{y}}^2 + C_2)}
\end{equation}
where $\mu _{\hat {\boldsymbol{y}}}$ and $\mu _{\boldsymbol{y}}$ ($\sigma^2_{\hat {\boldsymbol{y}}}$ and $\sigma^2_{\boldsymbol{y}}$) are the means (variances) of $\hat {\boldsymbol{y}}$ and ${\boldsymbol{y}}$, respectively, $\sigma _{\hat{\boldsymbol{y}}{\boldsymbol{y}}}$ is the covariance between $\hat {\boldsymbol{y}}$ and ${\boldsymbol{y}}$, and $C_1$ and $C_2$ are two constants that stabilize the division with a weak denominator.

\begin{figure*}[!t]
\centering
\includegraphics[width=2\columnwidth, angle=0]{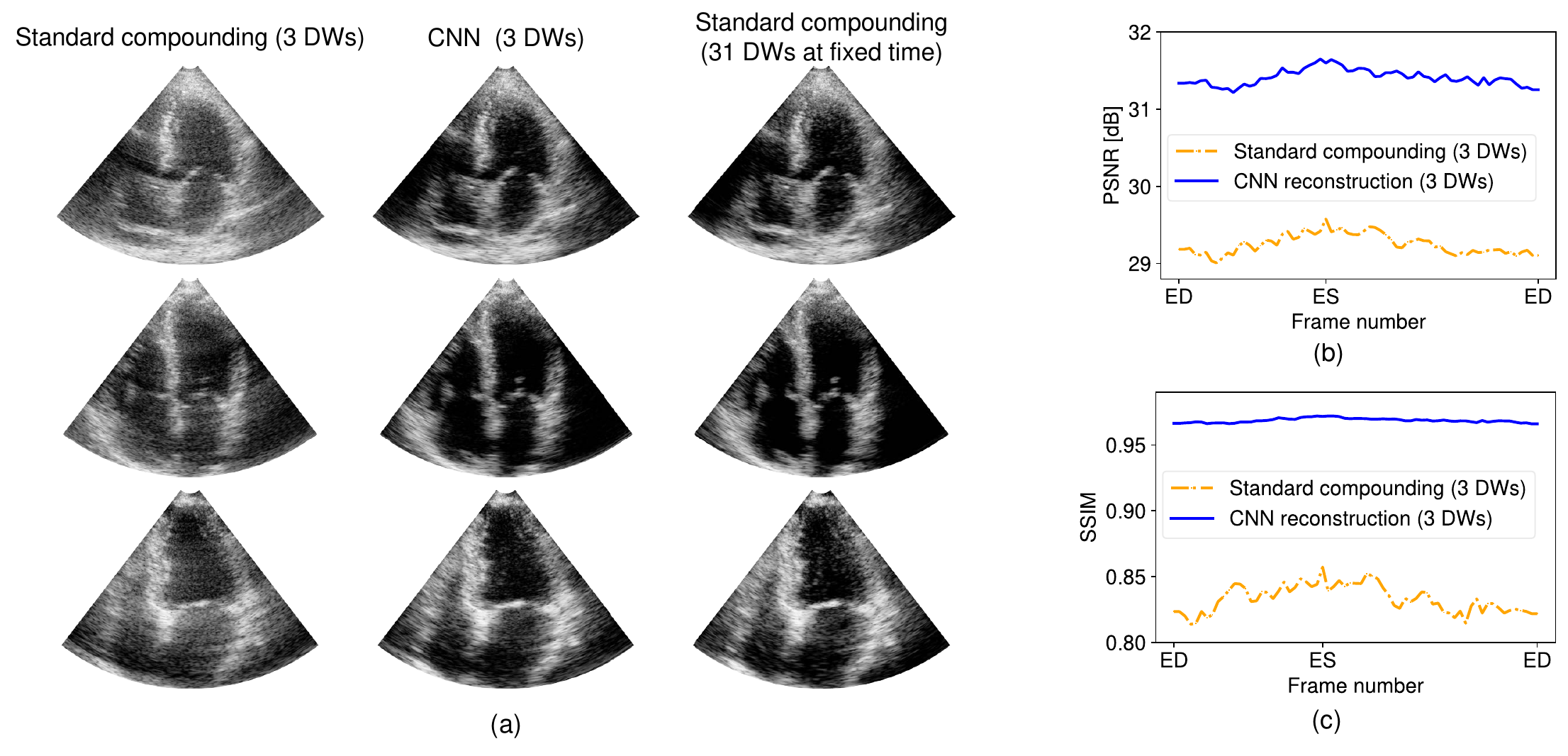}
\caption{(a) Typical reconstructions using standard compounding and CNN reconstruction from three DWs, along with the artifact-free reference reconstruction (from left to right) for 3 examples of cardiac simulations (from top to bottom). The images are displayed with a dynamic of 60 dB. (b) Average PSNR and (c) SSIM obtained using standard compounding and CNN reconstruction from three DWs for one cardiac cycle. For each time point, the average was calculated over the entire test set.}
\label{image_metric_simu}
\end{figure*}

For the \textit{in vitro} experiments, we measured the contrast-to-noise ratio (CNR) and generalized contrast-to-noise ratio (gCNR) of the anechoic cysts in the disk phantom. 
\begin{equation}
\label{cnr_formula}
{\rm CNR} = 20 \log_{10}\frac{\vert \mu _c - \mu _b\vert}{\sqrt{\sigma^2_c + \sigma^2_b}}
\end{equation}
where $\mu _c$ and $\mu _b$ ($\sigma^2_c$ and $\sigma^2_b$) are the means (variances) of the intensities within the cyst and the background regions, respectively. 

The gCNR is defined as the non-overlapping proportion between the intensity distribution within two regions.
\begin{equation}
\label{gcnr_formula}
{\rm gCNR} = 1- \sum_k min\{p_c(k), p_b(k)\}
\end{equation}
where $k$ is the pixel intensity, and $p_c(k)$ and $p_b(k)$ are the probability mass functions of the pixel intensity $k$ within the cyst and background regions.

\subsubsection{Motion estimation}
The accuracy of the estimated interframe motion was evaluated with the endpoint error (EPE), mean endpoint error (MEPE), and relative angular velocity error (RAVE). Given an estimated motion vector $\boldsymbol{\hat m}$ and the true motion vector $\boldsymbol{m}$, the EPE is computed as the Euclidean distance between $\boldsymbol{\hat m}$ and $\boldsymbol{m}$,
\begin{equation}
\mathrm{EPE}=\|\hat{\boldsymbol{m}}-\boldsymbol{m}\|_{2}
\end{equation}
For an ROI including $n$ measure points, the MEPE is computed as the average of all EPE values over the set.
\begin{equation}
\mathrm{MEPE}= \frac{1}{n} \sum _{i=1}^n  \|\hat{\boldsymbol{m}}_i-\boldsymbol{m}_i\|_{2}
\end{equation}
The RAVE was specifically used for the \textit{in vitro} experiment with the spinning disk.
\begin{equation}
\mathrm{RAVE}=\frac{\vert \hat{\omega} -\omega\vert}{\omega}
\end{equation}
where $\omega$ is the true angular velocity controlled by the motor, $\hat{\omega}$ is the estimated angular velocity, which was computed as the average of angular velocity at each inspected point within ROI covered by the spinning disk.

\section{Results}
\label{sec:results}

\subsection{Numerical Experiment}

\subsubsection{Image quality}

Fig. \ref{image_metric_simu}(a) displays three examples of reconstructed cardiac images in the numerical experiments. One can observe that the CNN improved the contrast and enhanced the myocardium textures, as compared with the images obtained from the standard compounding of the same three DWs. In particular, the images produced by the CNN are visually close to the reference images, i.e. motion artifact-free images generated with 31 steered DWs fired at a fixed time. 

We show in Fig.  \ref{image_metric_simu}(b) and \ref{image_metric_simu}(c) the average PSNR and SSIM obtained using standard compounding and CNN reconstruction from three DWs. The results are given for one cardiac cycle, by synchronizing the sequences to their end-diastolic (ED) and end-systolic (ES) frames. For each time point, the average was calculated over the entire test set.
From Fig. \ref{image_metric_simu} (b) and \ref{image_metric_simu} (c), 
the time variation of PSNR and SSIM associated with the two approaches showed a similar tendency: the values increased from ED to ES frame and then decreased from ES to ED frame. 
The CNN reconstruction showed less fluctuation along the cardiac cycle (average PSNR and SSIM), as compared to the standard compounding. This is particularly true for the SSIM results.

Table \ref{result_cardiac} reports the average PSNR and SSIM, as well as their standard deviations on all testing images. The standard compounding reached 29.28 dB in PSNR and 0.84 in SSIM. The CNN reconstruction reached 31.44 dB in PSNR and 0.97 in SSIM, which implies that CNN reconstructions are closer to the references, i.e. motion artifact-free reference images.

\begin{table}[!t]
\renewcommand{\arraystretch}{1.4}
\setlength{\tabcolsep}{0.2mm}
\caption{PSNR, SSIM, and MEPE reached by different approaches on the testing data.}
\label{result_cardiac}
\centering
\begin{tabular}{c c c c}
\toprule
Model & PSNR [dB]($\uparrow$) &  SSIM($\uparrow$) & MEPE [mm]($\downarrow$) \\ 
\midrule
Standard compounding (3 DWs)  & $29.28 \pm 0.56$  & $0.84 \pm 0.05$   & $0.23 \pm 0.03$\\
CNN reconstruction (3 DWs)  & $31.44\pm 0.50$ & $0.97\pm0.01$  & $0.15 \pm 0.02$\\
Standard compounding   & \multirow{2} * {–} & \multirow{2} * {–} & \multirow{2} * {$0.13\pm 0.01$}\\
\specialrule{0em}{-1pt}{-2pt}
(31 DWs at fixed time)\\
\bottomrule
\end{tabular}
\end{table}

\begin{figure*}[!t]
\centering
\includegraphics[width=2\columnwidth, angle=0]{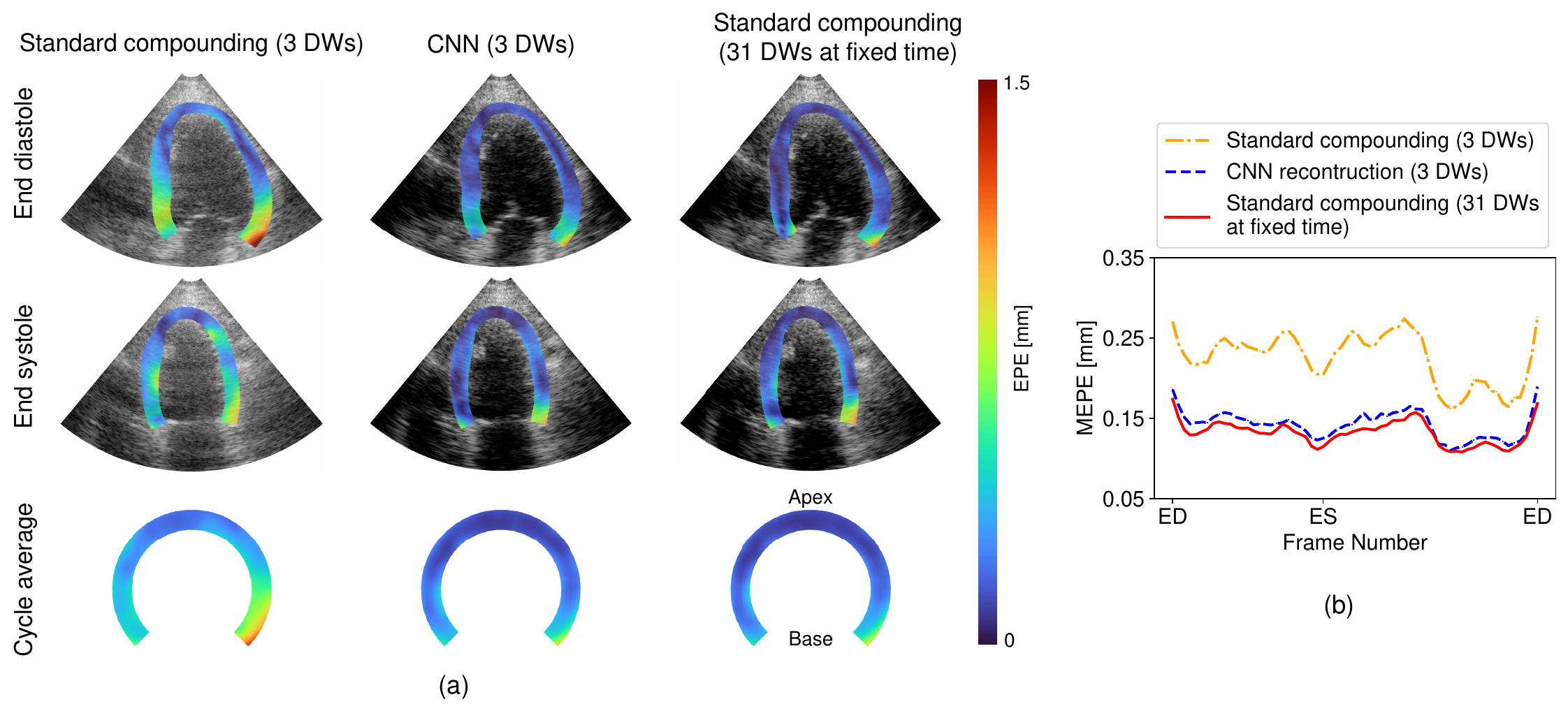}
\caption{(a) EPE spatial distribution of motion estimated from reconstructions using standard compounding and CNN reconstruction from three DWs, along with the result from the artifact-free reference (from left to right). Top and middle rows: typical EPE spatial distribution obtained for one test case at ED and ES. Bottom row: spatial distribution of the average EPE over the full cardiac cycle mapped on a common circular reference map for the same case. (b) Average EPE obtained using standard compounding, CNN reconstruction, and the motion artifact-free reference for one cardiac cycle. For each time point, the average was calculated over the entire test set.}
\label{image_metric_simu_motion}
\end{figure*}

\subsubsection{Motion estimation}

Fig. \ref{image_metric_simu_motion}(a) shows a typical EPE spatial distribution obtained for one test case at ED and ES, as well as time-average distribution (averaged over one cardiac cycle). 
Due to the varying shape of the left ventricle along the cardiac cycle, the spatial distribution of the averaged EPE was mapped on a circular reference map. 
It appears that the CNN reconstruction leads an EPE distribution close to the one computed from the reference, and lower than the one associated with the standard compounding of 3 DWs. 

Fig. \ref{image_metric_simu_motion}(b) shows the time variation of the MEPE for each approach. For each time point, the MEPE average was calculated over the entire test set.
One can observe that the three approaches showed a similar MEPE evolution tendency, while the CNN yielded a MEPE close to the one associated with the reference, and lower than the one associated with the standard compounding of 3 DWs throughout the whole cycle. Table \ref{result_cardiac} gives the MEPE for the three approaches, which were obtained by averaging the results on all testing data and time. Consistent with the observation in Fig. \ref{image_metric_simu_motion}(b), the CNN reconstruction and reference reached rather close results, which were 0.15 mm and 0.13 mm, respectively. The standard compounding of 3 DWs yielded a higher MEPE, i.e. 0.23 mm. 

\subsection{In Vitro Experiment}

\subsubsection{Image quality}

\begin{figure*}[!t]
\centering
\includegraphics[width=1.8\columnwidth, angle=0]{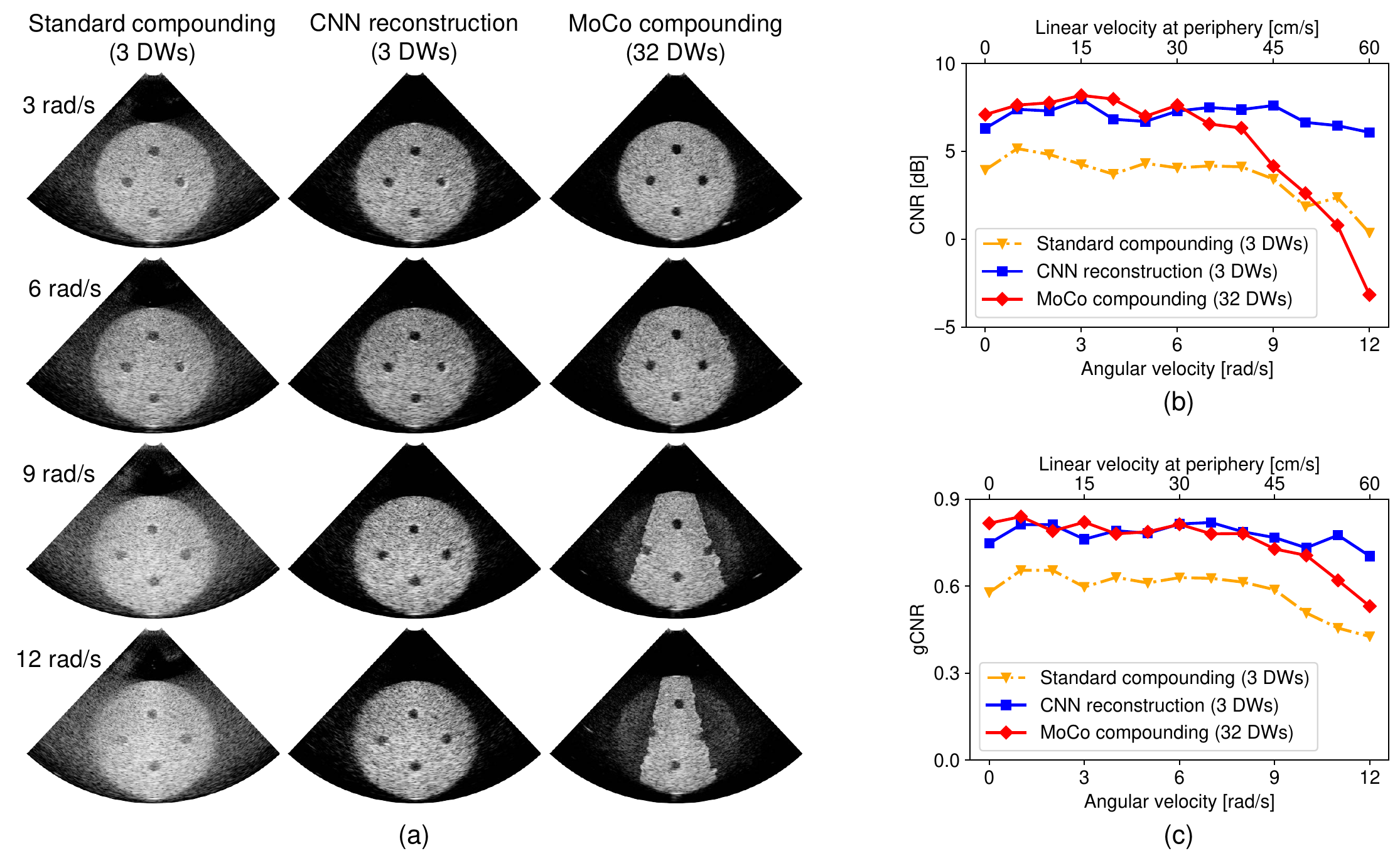}
\caption{(a) Reconstructed disk images at different rotational speeds, displayed in B-mode with a dynamic range of 60 dB. (b) CNR and (c) gCNR reached by different approaches at different disk rotational speeds. The CNR and gCNR were measured on the four anechoic cysts of the disk phantom at each rotational speed corresponding to the frames in (a).}
\label{image_metric_vitro}
\end{figure*}

\begin{figure*}[!t]
\centering
\includegraphics[width=1.9\columnwidth, angle=0]{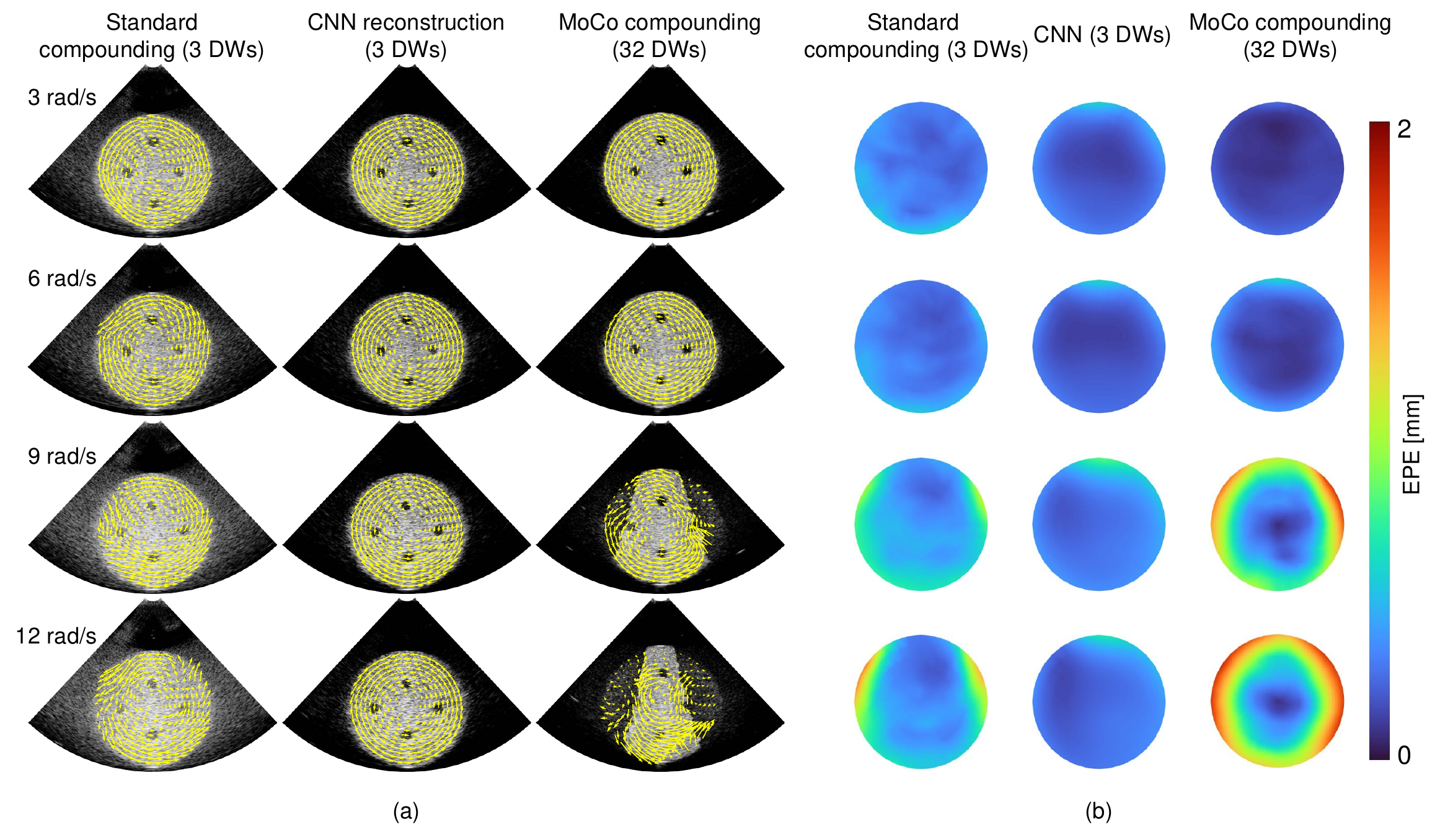}
\caption{(a) Motion fields and (b) EPE distribution of estimated motion on reconstructed disk sequences. Left to right: standard compounding of 3 DWs, CNN reconstruction of 3 DWs, and MoCo-based compounding of 32 DWs.}
\label{image_phantom_motion}
\end{figure*}

Fig. \ref{image_metric_vitro}(a) shows representative frames of reconstructed disk images in the \textit{in vitro} experiment. 
From Fig. \ref{image_metric_vitro}(a), it appears that the images obtained from the standard compounding of three DWs suffered from image artifacts, particularly for higher rotational speeds. 
Using the same three DW transmissions, the proposed CNN-based method restrained these artifacts, yielding a clear disk structure and enhanced contrast in the cyst regions. 
The MoCo-based imaging approach also reduced image artifacts. As the disk rotational speed increased, the MoCo-based approach failed to recover the disk structure in the high-speed regions. The reconstruction started to exhibit impaired structure on the disk rim at a rotational speed of 6 rad/s, corresponding to a rim linear speed of 30 cm/s. 

We evaluated the quality of the reconstructed images in terms of CNR and gCNR, as shown in Fig. \ref{image_metric_vitro}(b) and \ref{image_metric_vitro}(c). 
The results were measured on the four anechoic cysts of the disk phantom. 
As the contrast varies with the position in the field of view, the investigated images for each approach at each rotational speed correspond to the frames where the phantom cysts are in the “cross” position, as shown in Fig. \ref{image_metric_vitro}(a).
With the disk rotational speed below 6 rad/s, the CNN-based imaging with 3 DWs obtained a CNR of $\sim\,$7.5 dB level and a gCNR of $\sim\,$0.8 level, which was close to MoCo-based imaging with 32 DWs and higher than those obtained by the standard compounding of 3 DWs ($\sim\,$4.5 dB in CNR and $\sim\,$0.6 in gCNR).
With the rotational speed exceeding 6 rad/s, the CNR and gCNR reached by MoCo-based imaging decreased rapidly and dropped to its minimum of $-$3.2 dB in CNR and 0.53 in gCNR. 
Similarly, the results associated with the standard compounding decreased with higher disk speeds, reaching its minimum of 0.36 dB in CNR and 0.43 in gCNR. 
The proposed CNN-based imaging yielded stable results (though slightly decreased) as the rotational speed increased, and recovered a CNR of 6.6 dB and a gCNR of 0.72 at the highest rotational speed, resulting in a growing lead over the MoCo-based imaging and standard compounding.
These results are consistent with the visual observation in Fig. \ref{image_metric_vitro}(a)

\begin{figure}[!t]
\centering
\subfigure[]{\includegraphics[width=0.9\columnwidth]{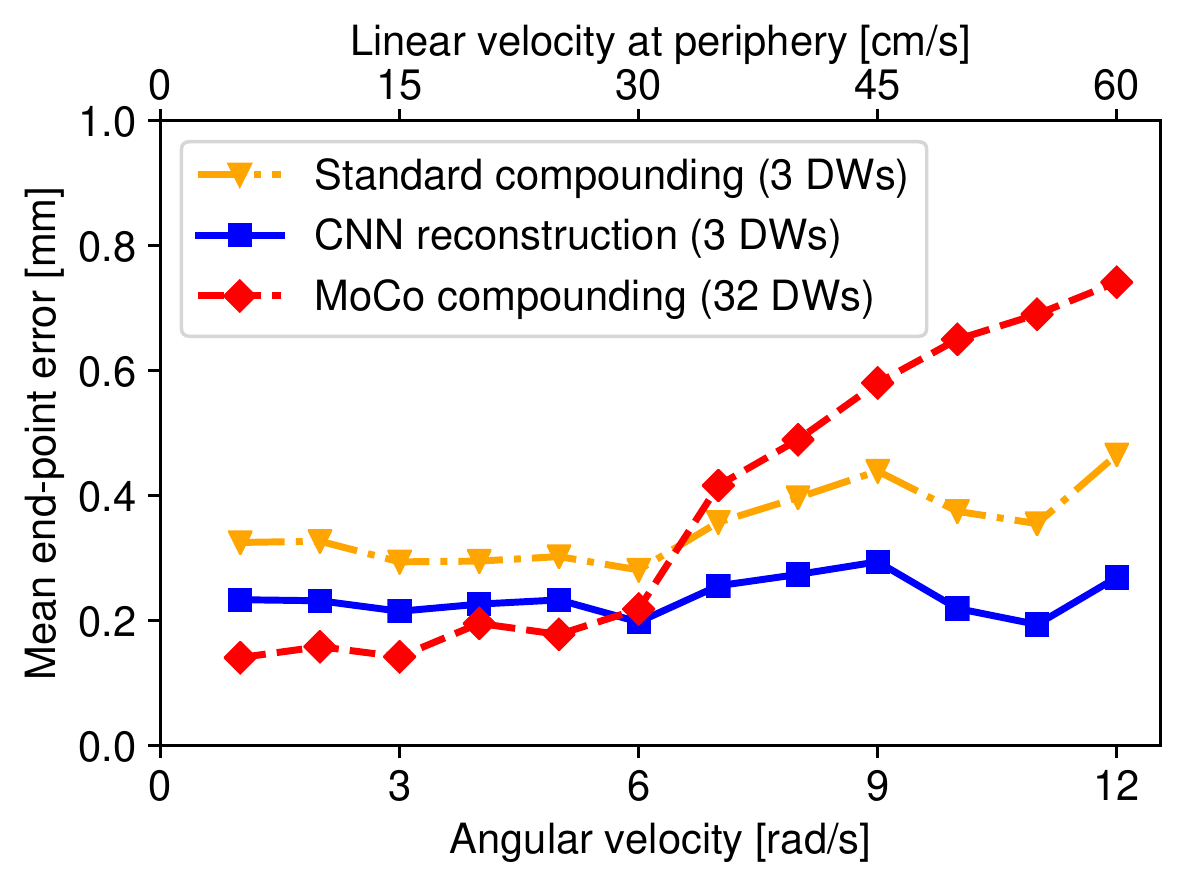}
\label{metric_mepe}}
\subfigure[]{\includegraphics[width=0.9\columnwidth]{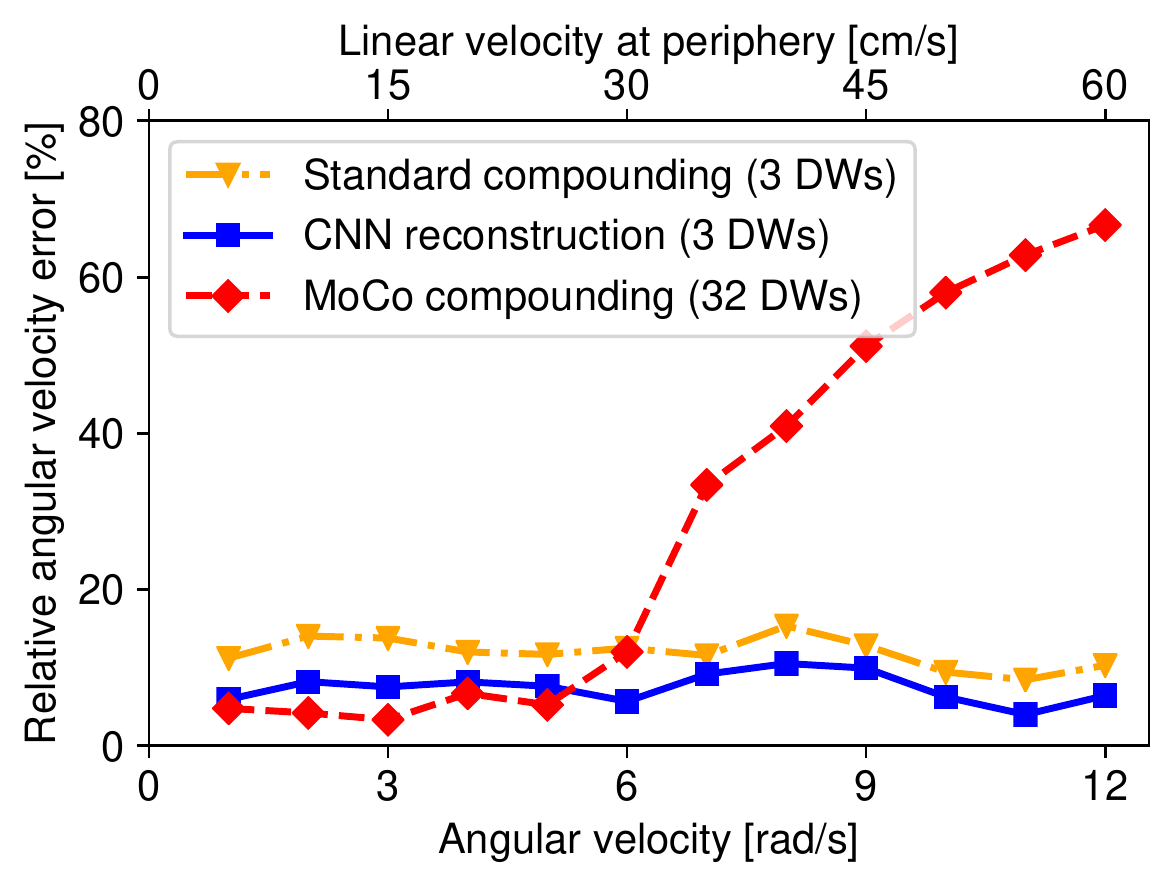}
\label{metric_rave}}
\caption{(a) MEPE and (b) RAVE reached by different approaches at different disk rotational speeds.}
\label{metric_mepe_rave}
\end{figure}

\subsubsection{Motion estimation}
 
Fig. \ref{image_phantom_motion} shows the motion field estimated by the speckle tracking and the EPE distribution of the estimated motion. One can observe that the standard compounding-based tracking suffered from the motion artifacts in the reconstructed images, exhibiting disturbed motion estimation as the disk rotational speed increased.
This is further demonstrated by the EPE distribution, particularly in the lateral region where the disk periphery was blurred due to the motion artifacts. 
Using the same three DW acquisitions, the proposed CNN-based approach provided a homogenous motion field in the ROI and yielded lower EPE for each disk rotational speed. 
In the same way, homogenous and accurate motion fields can be observed in the MoCo-based tracking with the disk spinning at no higher than 6 rad/s. As the disk rotational speed increased, the MoCo-based tracking failed to obtain homogenous motion fields due to the inconsistent speckle pattern in the impaired structure, yielding higher EPE in these regions.

Fig. \ref{metric_mepe_rave} reports the MEPE and RAVE associated with each approach at each rotational speed.  
From Fig. \ref{metric_mepe}, with the rotational speed below 6 rad/s, the MoCo-based tracking returned a MEPE below 0.20 mm, which was lower than the other approaches. A significantly increased MEPE was returned by the MoCo-based tracking as the rotational speed increased, reaching 0.74 mm at the highest rotational speed. 
The proposed CNN-based approach led a consistent MEPE in the range of 0.20 to 0.27 mm, which was lower than the standard compounding-based tracking (0.28 to 0.47 mm), regardless of the disk rotational speed.
In the same way, the proposed CNN-based approach returned a consistent RAVE ranging from 4\% to 10\% under a large range of rotational speed, which was lower than the standard compounding-based tracking (8\% to 15\%). The MoCo-based tracking led the lowest RAVE of $\sim\,$5\% with the rotational speed below 6 rad/s, while reaching up to 68\% as the rotational speed increased.

\subsection{In Vivo Experiment}
\begin{figure*}[!t]
\centering
\includegraphics[width=1.9\columnwidth, angle=0]{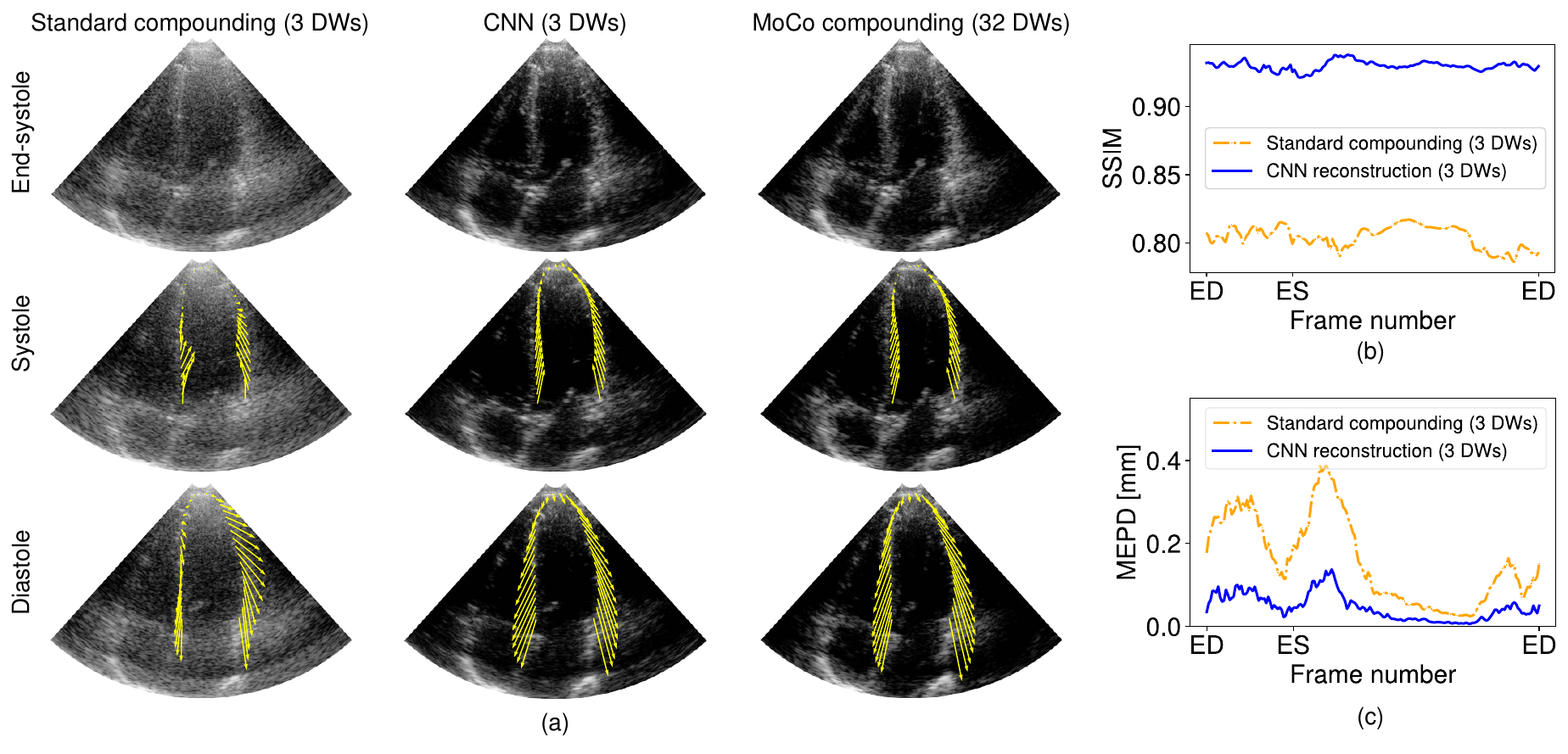}
\caption{(a) Reconstructed images at end-systole and left ventricular motion fields during systole and diastole for one example of \textit{in vivo} case (from top to bottom), obtained from standard compounding and CNN reconstruction using three DWs, along with MoCo-based compounding from 32 DWs (from left to right). (b) Average SSIM and (c) Mean endpoint difference obtained using standard compounding and CNN reconstruction from three DWs for one cardiac cycle. For each time point, the average was calculated over the five \textit{in vivo} cases.}
\label{image_metric_vivo}
\end{figure*}

Fig. \ref{image_metric_vivo} (a) displays one example of reconstructed \textit{in vivo} cardiac images at end-systole (first row) and left ventricular motion fields during systole (second row) and diastole (third row).
It appears that CNN-based reconstruction improved the contrast and enhanced anatomical structures as compared to the standard compounding of the same three DWs, and yielded an image quality close to the MoCo-based reconstruction. 
Furthermore, the motion fields associated with CNN-based and MoCo-based tracking showed very similar directions and amplitudes, while standard compounding-based tracking failed to provide homogenous motion vectors.

For each frame, the image quality and motion estimation associated to the 3-angle-based imaging (i.e. standard compounding and CNN-based reconstruction) were compared to the results provided by the 32-angle MoCo-based imaging (considered as the reference) using SSIM and mean endpoint difference (MEPD).
As opposed to the EPE measure, the MEPD is only a comparison measure, since the true cardiac motion is unknown. The equation of the MEPD is thus the following:
\begin{equation}
\mathrm{MEPD}= \frac{1}{n} \sum _{i=1}^n  \|\hat{\boldsymbol{m}}^{\mathrm{CNN}}_i-\hat{\boldsymbol{m}}^{\mathrm{MoCo}}_i\|_{2}
\end{equation}
where $n$ is the number of measure points, $\hat{\boldsymbol{m}}^{\mathrm{CNN}}_i$ and $\hat{\boldsymbol{m}}^{\mathrm{MoCo}}_i$ are the motion vectors estimated from the CNN reconstruction and MoCo reconstruction at measure point $i$.

Fig. \ref{image_metric_vivo} (b) and (c) show the time variation of the SSIM and MEPD. 
For each time point, the results were averaged over the five \textit{in vivo} cases in one cardiac cycle. 
The CNN leads to a clearly higher SSIM than that associated with standard compounding, regardless of the time in the cardiac cycle. Regarding motion estimation, the CNN yields a lower MEPD (with a maximum of 0.12 mm) and less fluctuation than standard compounding. Consistently, the difference between the two approaches narrows considerably in late diastole, when the cardiac motion is small.

\section{Discussion}
\subsection{Learning Ultrafast Cardiac Imaging}

In this study, we introduced a methodology for learning ultrafast cardiac imaging.
The imaging process was formulated as a non-linear mapping between the input space of low-quality images from ultrafast DW acquisitions to the output space of high-quality cardiac images, which was solved by supervised learning with respect to high-quality motion artifact-free references.
Considering the dynamic nature of cardiac imaging, acquiring such reference is challenging, if not possible, under real conditions. We resolved this challenge by adopting a carefully-designed simulation pipeline.
 Taking advantage of numerical simulations, benchmark images can be pushed “beyond physics” by simulating optimal DW insonifications of a static heart medium to eliminate motion artifacts. 
Leveraging such a dataset, a complex-valued CNN underwent learning for seeking an optimal reconstruction operator, specifically trained to restore high-quality I/Q images from only three successive DW transmissions. 

Visual assessment of the simulated cardiac images (Fig. \ref{image_metric_simu} (a)) and the obtained metrics (Table \ref{result_cardiac}) demonstrated that the proposed approach significantly improved the image quality compared with standard compounding using the same three DW transmissions.  
\textit{In vitro} and \textit{in vivo} results (Fig. \ref{image_metric_vitro} and \ref{image_metric_vivo}) showed consistent performance of the proposed method under real conditions, though the model was trained solely in the subspace of simulated cardiac data.
Moreover, equivalent imaging quality was demonstrated in qualitative and quantitative assessments, as compared with motion artifact-free compounding in numerical experiments and MoCo-based compounding in physical experiments, which indicates a good generalization of the proposed approach under different imaging conditions.

Speckle tracking results in the numerical experiment associated with the proposed approach confirm the preservation of temporal consistency of tissue motion among successive reconstructed frames, with a MEPE very close to the reference throughout the cardiac cycle (only an increase of 0.02 mm in average MEPE over the full cardiac cycle). \textit{In vivo} cardiac results confirmed this trend, yielding a maximum mean endpoint difference of 0.12 mm between CNN and MoCo. This demonstrates that our CNN approach successfully learned an effective reconstruction operator for the underlying speckle pattern, toward a motion artifact-free quality. 
 
These results demonstrate the feasibility of applying the deep learning-based method in ultrafast cardiac imaging, not only for improving exhibited image quality in separated frames but also preserving consistent speckle patterns in consecutive frames, which enables analysis of cardiac dynamics at ultra-high frame rates.

\subsection{Imaging and Tracking on High-Velocity Tissues}

Assessment of cardiac dynamics by speckle tracking requires high-quality imaging with consistent speckle patterns, which can be challenging when imaging and tracking high-velocity tissues. 
Ultrasound imaging with unfocused wavefronts can provide ultrahigh frame rates, but suffers from a low image quality if the transmission number for one frame is limited. 
As expected, the limitation was demonstrated with the \textit{in vitro} experimental results associated with the standard compounding of 3 DWs, in terms of both imaging quality and speckle-tracking accuracy. 
Fig. \ref{image_metric_vitro} illustrates the poor visual quality of the reconstructed images from this approach. The quantitative assessment in terms of CNR and gCNR hold for this visual observation, yielding the lowest results (except for the CNR in the 11- and 12-rad/s cases which were higher than the MoCo-based method). 
Motion artifacts were observed in high-velocity cases from Fig. \ref{image_metric_vitro}(a), leading to poorer speckle tracking in Fig. \ref{image_phantom_motion} and increased MEPE in Fig. \ref{metric_mepe_rave}.

As high-quality DW imaging generally relies on coherently compounding multiple steered transmissions, motion compensation (MoCo) is required to deal with the inter-acquisition motion when imaging fast-moving tissues. 
The implemented MoCo-based method indeed demonstrated improved imaging quality in the \textit{in vitro} experiment, as shown in Fig. \ref{image_metric_vitro}(a) (0- and 3-rad/s cases), and yielded superior CNR and gCNR results when the rotational speed was not high. 
As the disk rotational speed increased above 6 rad/s, a performance deterioration in image quality and motion estimation was observed in Fig. \ref{image_metric_vitro}(a) and \ref{image_phantom_motion}. Such degradation can be demonstrated by the drop in CNR and gCNR, as well as the increase in MEPE and RAVE.
The maximum detectable velocity of effective MoCo is related to its Doppler Nyquist limit, which can be derived by $c\,PRF/(8\,f_0)$\cite{poree2016high} (with $f_0=$ carrier frequency). Considering a speed of sound $c$ of 1540 m/s and the imaging settings in the experiments, the deployed MoCo reached a threshold of $\sim\,$29 cm/s, which is consistent with the observation in the 6-rad/s case (corresponding to 30 cm/s at the disk rim). 
For lower speed cases, MoCo is fully operational, while the CNN compensates slightly less for the motion intervening between each steered DW acquisition.


The interest of the present study was to improve image quality while preserving the temporal consistency of high-velocity tissue motion, thereby pushing the limit of detectable velocity for high-quality ultrasound imaging and accurate speckle tracking.
An important observation from Fig. \ref{image_metric_vitro}(a) and \ref{image_phantom_motion} is that the proposed method provided robust imaging quality and enabled accurate motion estimation under a large range of rotational velocities. 
Fig. \ref{image_metric_vitro}(b), \ref{image_metric_vitro}(c), and \ref{metric_mepe_rave} further demonstrate steady performance in all evaluation metrics preserved by the proposed method, particularly at high velocities that exceed the limit of the MoCo-based imaging. 
Reliable speckle tracking was demonstrated in the 12-rad/s case, reaching the maximum velocity at the boundary with a value of 60 cm/s. Imaging modalities for higher velocities, such as stress echocardiography and blood flow tracking, may thus benefit from such detectable velocity. 


The \textit{in vivo} experiments aimed at exploring the feasibility and reliability of the proposed approach for future use in clinical settings, using a 32 DWs triangular MoCo sequence as a reference. The results (Fig. \ref{image_metric_vivo}) were obtained from normal subjects and indicate that the CNN approach is on par with the MoCo approach in terms of motion estimation (maximum mean endpoint difference of 0.12 mm), while offering the possibility of much higher frame rates (1500 frames/second when using a PRF of 4500 Hz).

\subsection{Perspectives}
The proposed approach provides a feasible solution for ultrafast speckle-tracking echocardiography, while relying on two separate steps, i.e. CNN-based image reconstruction from ultrafast DW transmissions and speckle tracking on reconstructed sequences.
This process may be refined in an end-to-end manner by adopting multi-task learning for both imaging reconstruction and motion tracking, as deep learning methods have also been developed for motion estimation\cite{evain2022motion, tehrani2020displacement,delaunay2021unsupervised,wei2022unsupervised,ostvik2021myocardial,tehrani2022lateral}. The model accuracy may be further improved by considering joint optimization for multiple tasks and possible motion compensation from the estimated motion. 
Moreover, as the proposed approach adopts data-driven deep learning, the training data is also a crucial factor for possible improvement. In the present study, the myocardial motions in the training data were limited to planar simulations. Integrating out-of-plane motions in the myocardium simulation may be interesting, as they generally occur in physical conditions and may bring additional speckle inconsistency in 2D. It would also be interesting to incorporate intracardiac flow (as recently proposed by \cite{sun2022pipeline}) in the training data for exploiting analysis diversity from the proposed ultrafast imaging, since DW-based echocardiography is well-adapted for simultaneous analysis of myocardial motion and intracardiac flow. 

\section{Conclusion}
In this article, a methodology for ultrafast cardiac imaging based on deep learning was presented, particularly implemented in conjunction with speckle tracking for investigating its temporal property for motion estimation.
The method consists of reconstructing high-quality cardiac images from ultrafast DW acquisitions by adopting a complex-valued CNN and supervised learning from a simulated echocardiographic dataset.
We evaluated the performance of the proposed method in numerical, \textit{in vitro}, and \textit{in vivo} experiments, and demonstrated its effectiveness in both imaging quality enhancement and speckle pattern restoration for reliable speckle tracking, under a large range of tissue velocities, while allowing high frame rates. As such, cardiac diagnosis using dynamic analysis can benefit from the high imaging quality, frame rate, and perceptible speed of the proposed approach. 
 
\bibliographystyle{IEEEtran}
\bibliography{Deep_Cardiac_Imaging.bib}

\end{document}